
\font\titlefont = cmr10 scaled\magstep 4
\font\sectionfont = cmr10

\font\teenyfont = cmr5

\magnification = 1200

\global\baselineskip = 1.2\baselineskip
\global\parskip = 4pt plus 0.3pt
\global\nulldelimiterspace = 0pt

\predisplaypenalty 1000


\def\endignore{}
\def\ignore #1\endignore{}

\newcount\dflag
\dflag = 0


\def\monthname{\ifcase\month
\or Jan \or Feb \or Mar \or Apr \or May \or June%
\or July \or Aug \or Sept \or Oct \or Nov \or Dec
\fi}




\def\endid{}
\def\id#1\endid{\number\day\ \monthname \number\year
\hfill #1}

\def\endtitle{}
\def\title#1\endtitle{\vskip.3in\titlefont
\global\baselineskip = 2\baselineskip
#1\vskip.4in
\baselineskip = 0.5\baselineskip\rm}

\def\lblfoot{This work was supported by the Director, Office of Energy
Research, Office of High Energy and Nuclear Physics, Division of High
Energy Physics of the U.S. Department of Energy under Contract
DE-AC03-76SF00098.}

\def\endauthors{}
\def\authors#1\endauthors{
#1\if\dflag = 0
\footnote{}{\noindent\lblfoot}\fi}

\def\endabstract{}
\def\abstract#1\endabstract{\vskip .3in%
\centerline{\sectionfont\bf Abstract}%
\vskip .1in%
\noindent#1%
\ifnum\dflag = 0
\footline = {\hfil}\pageno = 0
\vfill\eject
\pageno = 1\footline{\centerline{\sectionfont\folio}}
\fi\ifnum\dflag = 2
\footline = {\hfil}\pageno = 0
\vfill\eject
\fi}

\newcount\nsection
\newcount\nsubsection

\def\section#1{\global\advance\nsection by 1
\global\nsubsection = 0
\bigskip\centerline{\sectionfont\bf\number\nsection.\ #1}%
\nobreak\medskip\rm\nobreak}

\def\subsection#1{\global\advance\nsubsection by 1
\bigskip\centerline{\sectionfont\it\number\nsection.\number\nsubsection.\ #1}%
\nobreak\medskip\rm\nobreak}

\def\appendix#1#2{\bigskip\noindent%
\sectionfont \bf Appendix #1.\ #2
\nobreak\medskip\rm\nobreak}


\newcount\nref
\global\nref = 1

\def\ref#1#2{\xdef #1{[\number\nref]}
\ifnum\nref = 1\global\xdef\therefs{\noindent[\number\nref] #2\ }
\else
\global\xdef\oldrefs{\therefs}
\global\xdef\therefs{\oldrefs\vskip.1in\noindent[\number\nref] #2\ }%
\fi%
\global\advance\nref by 1
}

\def\listrefs{\vfill\eject\section{References}\therefs}


\newcount\nfig
\global\nfig = 1

\def\fg#1\efig{\vskip .5in\noindent Fig.\ \number\nfig:\ #1%
\global\advance\nfig by 1}


\newcount\cflag
\newcount\nequation
\global\nequation = 1
\def\eqlabel{(1)}

\def\nexteqno{\ifnum\cflag = 0
\global\advance\nequation by 1
\fi
\global\cflag = 0
\xdef\eqlabel{(\number\nequation)}}

\def\lasteqno{\global\advance\nequation by -1
\xdef\eqlabel{(\number\nequation)}}

\def\label#1{\xdef #1{(\number\nequation)}
\ifnum\dflag = 1
{\escapechar = -1
\xdef\draftname{\teenyfont\string#1}}
\fi}

\def\clabel#1#2{\xdef\eqlabel{(\number\nequation #2)}
\global\cflag = 1
\xdef #1{\eqlabel}
\ifnum\dflag = 1
{\escapechar = -1
\xdef\draftname{\string#1}}
\fi}

\def\cclabel#1#2{\xdef\eqlabel{#2)}
\global\cflag = 1
\xdef #1{\eqlabel}
\ifnum\dflag = 1
{\escapechar = -1
\xdef\draftname{\string#1}}
\fi}


\def\eeq{}

\def\eqnn #1\eeq{$$ #1 $$}

\def\eq #1\eeq{\xdef\draftname{\ }
$$ #1
\eqno{\eqlabel \rlap{\ \draftname}} $$
\nexteqno}







\def\eqa #1\eeq{\xdef\draftname{\ }
$$ \eqalignno{ #1 } $$
\global\cflag = 0}


\def\ie{{\it i.e.\/}}
\def\eg{{\it e.g.\/}}



\def\myinstitution{
    \centerline{\it Theoretical Physics Group}
    \centerline{\it Lawrence Berkeley Laboratory}
    \centerline{\it 1 Cyclotron Road}
    \centerline{\it Berkeley, California 94720}
}


\def\jref#1#2#3#4{{\it #1} {\bf #2}, #3 (#4)}

\def\HPA#1#2#3{\jref{Helv.\ Phys.\ Acta}{#1}{#2}{#3}}

\def\NPB#1#2#3{\jref{Nucl.\ Phys.}{B#1}{#2}{#3}}
\def\PA#1#2#3{\jref{Physica}{#1A}{#2}{#3}}
\def\PLB#1#2#3{\jref{Phys.\ Lett.}{#1B}{#2}{#3}}

\def\PRD#1#2#3{\jref{Phys.\ Rev.}{D#1}{#2}{#3}}

\def\PRL#1#2#3{\jref{Phys.\ Rev.\ Lett.}{#1}{#2}{#3}}
\def\PRV#1#2#3{\jref{Phys.\ Rev.}{#1}{#2}{#3}}


\def\goto{\mathop{\rightarrow}}


\def\myint{\int\mkern-5mu}
\def\frac#1#2{{{#1} \over {#2}}\,}  
\def\sfrac#1#2{{\textstyle\frac{#1}{#2}}}  

\def\part{\partial}

\def\Dsl{\hbox{/\kern-.6000em\rm D}} 



\def\mybar#1{\kern 0.8pt\overline{\kern -0.8pt#1\kern -0.8pt}\kern 0.8pt}

\def\sla#1{\raise.15ex\hbox{$/$}\kern-.57em #1}
\def\Sla#1{\kern.15em\raise.15ex\hbox{$/$}\kern-.72em #1}

\def\roughly#1{\mathrel{\raise.3ex\hbox{$#1$\kern-.75em%
    \lower1ex\hbox{$\sim$}}}}
\def\lsim{\roughly<}

\def\scr#1{{\cal #1}}


\def\al{\alpha}
\def\gam{\gamma}

\def\Del{\Delta}

\def\lam{\lambda}
\def\Lam{\Lambda}

\def\Sig{\Sigma}


\def\tr{\mathop{\rm tr}}


\def\bra#1{\langle #1 |}
\def\ket#1{| #1 \rangle}




\def\hc{{\rm h.c.}}


\def\MeV{{\rm \ MeV}}
\def\GeV{{\rm \ GeV}}

\hyphenation{ba-ry-on ba-ry-ons}

\def\chisim{$SU(3)_L \times SU(3)_R$}


\id
LBL-33435
\endid

\title
\centerline{Chiral Corrections to Hyperon}
\centerline{Vector Form Factors}
\endtitle

\authors
\centerline{Jeffrey Anderson\ \ {\it and}\ \ Markus A. Luty}
\footnote{}{\lblfoot}
\vskip .1in
\myinstitution
\endauthors

\abstract
We show that the leading $SU(3)$-breaking corrections to the $\Delta S = 1$
$f_1$ vector form factors of hyperons are $O(m_s)$ and $O(m_s^{3/2})$,
and are expected to be $\sim 20$--$30\%$ by dimensional analysis.
This is consistent with the Ademollo--Gatto theorem, in a sense that
we explain.
We compute the $O(m_s)$ corrections and a subset of the $O(m_s^{3/2})$
corrections using an effective lagrangian in which the baryons are
treated as heavy particles.
All of these corrections are surprisingly small, $\sim 5\%$;
combining them, we obtain $\sim 5$--$10\%$ corrections.
The pattern of corrections is very different than that predicted by
quark models.
\endabstract


\ref\AG{M. Ademollo and R. Gatto, \PRL{13}{264}{1964}.}

\ref\heavy{H. Georgi, \PLB{240}{447}{1990};
T. Mannel, W. Roberts, and Z. Ryzak, \NPB{368}{204}{1992}.}

\ref\JM{E. Jenkins and A. V. Manohar, \PLB{255}{558}{1991}.}

\ref\fit{E. Jenkins and A. V. Manohar, \PLB{259}{353}{1991}.}

\ref\Krause{A. Krause, \HPA{63}{3}{1990}.}

\ref\quarkmod{See \eg\ J. F. Donoghue and B. R. Holstein,
\PRD{25}{206}{1982}.}

\ref\effL{J. Schwinger, \PLB{24}{473}{1967};
S. Coleman, J. Wess, and B. Zumino, \PRV{117}{2239}{1969};
C. G. Callan, S. Coleman, J. Wess, and B. Zumino, \PRV{117}{2247}{1969};
S. Weinberg, \PA{96}{327}{1979}.}

\ref\GL{See \eg\ J. Gasser and H. Leutwyler, \NPB{250}{1985}{465}.}

\ref\oldfit{R. L. Jaffe and A. V. Manohar, \NPB{337}{509}{1990}.}

\ref\determ{J. F. Donoghue, B. R. Holstein, and S. W. Klimt,
\PRD{35}{934}{1987};
Particle Data Group, {\it Phys.\ Rev.} {\bf D41} Part II, (1992).}


\section{Introduction}

In this paper, we consider the application of chiral perturbation theory to
the $f_1$ vector form factor of octet baryon states.
The form factors of the vector current are conventionally defined by
\eq
\bra{B_a} J_{\mu c}(0) \ket{B_b} = \mybar u(p_a) \biggl[[
f_1^{abc}(q^2) \gamma_\mu
+ \frac{if_2^{abc}(q^2)}{M_a + M_b} \sigma_{\mu\nu} q^\nu
+ \frac{if_3^{abc}(q^2)}{M_a + M_b} q_\mu \biggr] u(p_b),
\eeq
where $q \equiv p_a - p_b$.
Our interest in the form factor $f_1$ is due to the fact that it is
usually assumed that $SU(3)$ breaking corrections to $f_1$ are small due
to the Ademollo--Gatto (AG) theorem \AG.
Indeed, nonrelativistic quark model and bag model calculations of $SU(3)$
breaking corrections to $f_1$ typically give corrections of order
$1\%$ \quarkmod.

In this paper, we point out that the leading corrections to $f_1$ due to the
nonvanishing strange quark mass are $O(m_s)$ and $O(m_s^{3/2})$.
The $O(m_s)$ terms are proportional to
\eq
\frac{m_K^2}{16 \pi^2 f^2} \sim 0.2,
\eeq
where $f \simeq 93 \MeV$.
The $O(m_s^{3/2})$ corrections consist of terms proportional to
\eq
\frac{m_K \Delta_B}{16\pi f^2} \sim 0.2, \qquad
\frac{m_K^3}{16\pi f^2 \Lambda} \sim 0.3,
\eeq
where $\Lam \sim 1 \GeV$ is the expansion scale in chiral perturbation
theory, and $\Delta_B$ is an octet baryon mass splitting.
Clearly, it is important to compute these corrections, since dimensional
analysis does not guarantee that they are small.
For example, they could affect the determination of $D$ and $F$ from
semileptonic hyperon decay rates.
(In the formalism we employ, $D$ and $F$ are defined as couplings
in an effective lagrangian which embodies the low-energy theorems for
chiral symmetry.
In the $SU(3)$ limit, we recover well-known relations such as
$D + F = g_A$, but there are $SU(3)$-breaking corrections to these
relations due to nonvanishing quark masses which can be substantial.)

We compute the $O(m_s)$ corrections using an effective lagrangian in
which the baryons are treated as heavy particles \heavy\JM.
Using the values $D = 0.61$, $F = 0.40$ determined using a recent fit to
semileptonic hyperon decay \fit, we find these corrections to be surprisingly
small, $\lsim 5\%$.
These corrections have been computed in ref.\ \Krause, and we agree with
the results of this paper.
The $O(m_s^{3/2})$ contributions proportional to $m_K \Delta_B$ have not
been previously computed.
We find that these corrections are also $\lsim 5\%$ for all decays.
The terms proportional to $m_K^3$ cannot be computed in terms of the
lowest order chiral lagrangian.\footnote{$^\dagger$}
{Ref.\ \Krause\ gives some $O(m_s^{3/2})$ corrections of the form
$m_K^3 / (16\pi f^2 M_B)$ where $M_B$ is the average octet baryon mass.
These contributions are an artifact of the method of computation used
in that paper, and including them is not justified.}
The predicted corrections are increased significantly if the lowest-order
fit values of $D$ and $F$ are used, or if we make the approximation
$m_\pi \simeq 0$.
Our computation gives some indication that the corrections to $f_1$ are
$\lsim 10\%$, but we cannot exclude the possibility that the remaining
$O(m_s^{3/2})$ corrections are $\sim 30\%$ or more.

The plan of this paper is as follows:
In section 2, we review the effective lagrangian formalism we use to carry
out the computation.
In section 3, we discuss the Ademollo--Gatto theorem and how it is
manifested in the effective lagrangian framework.
The reader eager for the bottom line can skip immediately to
section 4, in which we present our results.
Section 5 contains our conclusions.

\section{The Effective Lagrangian}

It has been known for some time that the low-energy theorems of chiral
symmetry breaking are {\it equivalent} to a description of the low-energy
dynamics in terms of an effective lagrangian \effL.
Recently, it was realized that baryons could be simply included in an
effective lagrangian framework using an heavy particle effective theory
\heavy\JM.
This approach provides significant conceptual and calculational advantages:
the non-relativistic limit is incorporated from the start, and the Feynman
rules for computing graphs are considerably simplified.

In this section, we define the effective lagrangian and establish our
notation.

\subsection{Mesons}

The field
\eq
\xi(x) = e^{i\Pi(x) / f},
\eeq
is taken to transform under $SU(3)_L \times SU(3)_R$ as
\eq
\xi \mapsto L \xi U^\dagger = U \xi R^\dagger,
\eeq
where this equation implicitly defines $U$ as a function of $L$, $R$,
and $\xi$.
The meson fields are
\eq
\Pi = \frac 1{\sqrt 2}
\pmatrix{\frac 1{\sqrt 2}\pi^0 + \frac 1{\sqrt 6}\eta &
\pi^+ & K^+ \cr
\pi^- & -\frac 1{\sqrt 2} \pi^0 + \frac 1{\sqrt 6}\eta &
K^0 \cr
K^- & {\mybar K}^0 & -\frac 2{\sqrt 6} \eta \cr}.
\eeq

We will be interested in matrix elements of the vector currents.
We therefore add to the effective lagrangian a source term
\eq
\delta\scr L = \scr V_\mu J^\mu_V + \scr A_\mu J^\mu_A,
\eeq
where $J^\mu_V$ ($J^\mu_A$) is the vector (axial vector) Noether
current.
The couplings of $\scr V_\mu$ and $\scr A_\mu$ in the effective
lagrangian are then determined by demanding that they transform as
gauge fields (see eq.\ (12)).
We therefore define the covariant derivatives
\eq
D_\mu \xi \equiv \partial_\mu \xi - i\scr L_\mu \xi,
\qquad D_\mu \xi^\dagger \equiv \partial_\mu \xi^\dagger
- i\scr R_\mu \xi^\dagger.
\eeq
(Note that $(D_\mu\xi)^\dagger \ne D_\mu\xi^\dagger$.)
Here,
\eq
\scr R_\mu = \frac 12\left(\scr V_\mu + \scr A_\mu\right),
\qquad \scr L_\mu = \frac 12\left(\scr V_\mu - \scr A_\mu\right).
\eeq
$\scr V_\mu$ and $\scr A_\mu$ are hermitian.
The effective lagrangian is most conveniently written in terms of
\eq
V_\mu \equiv \frac i2\left(\xi D_\mu \xi^\dagger
+ \xi^\dagger D_\mu \xi\right), \qquad
A_\mu \equiv \frac i2\left(\xi D_\mu \xi^\dagger
-\xi^\dagger D_\mu \xi\right),
\eeq
which transform under {\it local} \chisim\ transformations as
\eq
V_\mu \mapsto U V_\mu U^\dagger + iU\partial_\mu U^\dagger, \qquad
A_\mu \mapsto U A_\mu U^\dagger,
\eeq
since the sources transform as gauge fields:
\eq
\scr L_\mu \mapsto L\scr L_\mu L^\dagger + iL\partial_\mu L^\dagger, \qquad
\scr R_\mu \mapsto R\scr R_\mu R^\dagger + iR\partial_\mu R^\dagger.
\eeq
Note that $A_\mu$ and $V_\mu$ are hermitian.
We can then define the covariant derivative
\eq
\label\covder
\nabla_\mu A_\nu \equiv \partial_\mu A_\nu - i [V_\mu, A_\nu],
\eeq
which transforms under local \chisim\ transformations as
\eq
\nabla_\mu A_\nu \mapsto U\nabla_\mu A_\nu U^\dagger.
\eeq

The chiral symmetry is broken explicitly by the quark masses.
(We neglect the effects of electromagnetism in this paper.)
We will ignore isospin breaking, so that the quark mass matrix is taken to be
\eq
\label\massmat
M_q = \pmatrix{\hat m &&\cr &\hat m&\cr &&m_s\cr}.
\eeq
It is convenient to define
\eq
M \equiv \frac 12\!\left( \xi^\dagger M_q \xi^\dagger + \hc\right)
\mapsto U\!M U^\dagger.
\eeq

The simple transformation rules of the fields defined above makes it easy
to write down the effective lagrangian.
For example, the leading terms can be written
\eq
\scr L_0 = f^2 \tr(A^\mu A_\mu) + af^3 \tr M.
\eeq

\subsection{Baryons}

We now discuss the inclusion of baryon fields as heavy particles \heavy\JM.
The momentum of a baryon field is written
\eq
P = M_B v + p,
\eeq
where $M_B$ is a $SU(3)$-invariant baryon mass and $v$ is a velocity.
The key observation is that for processes involving emission of soft pions,
the relevant residual momenta $p$ are small if $v$ is chosen appropriately.
For a fixed $v$, we can then write an effective theory in terms of baryon
fields $B$ whose momentum is the residual momentum $p$ of the baryon \heavy.

The octet baryon fields $B$ transform under \chisim\ as
\eq
B \mapsto U\!B U^\dagger.
\eeq
Explicitly, we have
\eq
B = \pmatrix{
\frac 1{\sqrt 2} \Sigma^0 + \frac 1{\sqrt 6}\Lambda &
\Sigma^+ & p \cr
\Sigma^- &
-\frac 1{\sqrt 2}\Sigma^0 + \frac 1{\sqrt 2}\Lambda & n \cr
\Xi^- & \Xi^0 & -\frac 2{\sqrt 6}\Lambda \cr}.
\eeq

The lowest order terms in the effective lagrangian involving baryon fields
are
\eq
\eqalign{
\scr L &= \tr\left(\mybar B iv\cdot\nabla B\right)
+ 2D \tr\left( \mybar B s^\mu \{ A_\mu, B \} \right)
+ 2F \tr\left( \mybar B s^\mu [ A_\mu, B ] \right) \cr
&\qquad + \sigma \tr\left(M \right) \tr\left( \mybar B B \right)
+ b_D \tr\left( \mybar B \{ M, B \} \right)
+ b_F \tr\left( \mybar B [ M, B ] \right), \cr}
\eeq
where $s^\mu$ is the spin operator \JM\ and the covariant derivative acts
on $B$ as in eq.\ \covder.

\subsection{Power Counting}

The effective lagrangian described above has a well-defined expansion
in inverse powers of $\Lambda \sim 1 \GeV$.
A typical term in the lagrangian can be written schematically
\eq
\scr L \sim f^2 \Lam^2 \left( \frac B{f\sqrt\Lam} \right)^{n_B}\!\!
\left( \frac\nabla\Lam \right)^{n_D}\!\!
\left( \frac A\Lam \right)^{n_A}\!\!
\left( \frac M\Lam \right)^{n_M}.
\eeq
If we write
\eq
m_\Pi^2 \sim a f M_q,
\qquad\qquad \Del_B \sim b M_q,
\eeq
then topological identities can be used to show that loop corrections are
related to tree-level contributions by
\eq
{\rm loop} \sim \left( \frac\Lam{4\pi f} \right)^{2L}
\left( \frac{af}\Lam \right)^{N_m / 2}
b^{N_\Del} \times {\rm tree},
\eeq
where $L$ is the number of loops in the diagram, and $N_m$ ($N_\Del$)
is the number of powers of $m_\Pi$ ($\Del_B$) in the result of the loop
diagram.
This expansion is consistent provided that $\Lam \lsim 4\pi f$,
$a \lsim \Lam / f \lsim 4\pi$, and $b \lsim 1$.
This appears to be satisfied in QCD \GL.

\section{The Ademollo--Gatto Theorem}

In this section, we review the Ademollo--Gatto (AG) theorem \AG\ and discuss
how it is realized in the effective lagrangian approach.
Much of this section is quite elementary, but we feel that the issues involved
deserve a careful treatment.

Suppose that a quantum-mechanical system has a global symmetry $G$ which is
explicitly broken by perturbations whose size is controlled by a parameter
$\lam$.
We assume $\lam$ is sufficiently small so that the explicit symmetry breaking
can be treated perturbatively.
It is then convenient to expand the physical states of the system in terms
of states with definite transformation properties under $G$:
\eq
\label\stadec
\ket\al = c_\al \ket{r_\al\, j_\al}
+ \sum_{r, j} c_{\al j}^r \ket{r \, j},
\qquad\qquad c_{\al j_\al}^{r_\al} \equiv 0.
\eeq
Here, $\ket{r \, j}$ is a state belonging to the irreducible representation
$r$; $j$ labels the particular state.
The state $\ket{r_\al\, j_\al}$ is the state corresponding to the
physical state $\ket\al$ in the limit $\lambda \goto 0$:
\eq
c_\al \goto 1, \qquad
c_{\al j}^r = O(\lambda), \qquad
{\rm as\ } \lambda \goto 0.
\eeq

The AG theorem applies  if the symmetry breaking effect is such that it
does not mix states from the same irreducible representation, \ie
\eq
\label\assum
c_{\al j}^{r_\alpha} = 0.
\eeq
In this case, the AG theorem states that for any charge $Q$ of $G$,
\eq
\label\AG
\bra\beta Q \ket\al = q_\al \delta_{\al\beta} + O(\lam^2),
\eeq
where $q_\al$ is the charge of the unperturbed state:
\eq
Q\ket{r_\al\, j_\al} = q_\al \ket{r_\al\, j_\al}.
\eeq
This theorem can be applied to the $f_1$ form factor, since
\eq
\label\formeq
\bra{B_a} Q_c \ket{B_b} = \myint d^3 x\, \bra{B_a} J_{0c}(x) \ket{B_b}
= u^\dagger(p_a) u(p_b) f_1^{abc}(\vec q = 0) + O(M_q^2).
\eeq
The conditions of the theorem are satisfied in the case of explicit
$SU(3)$ breaking due to the strange quark mass, since the mass matrix
eq.\ \massmat\ has definite isospin and hypercharge, and
thus does not mix members of the octet.

The proof of the AG theorem is by direct computation:
\eq
\bra\beta Q \ket\al =
c_\al c_\beta \bra{r_\beta\, j_\beta} Q \ket{r_\al\, j_\al}
+ \sum_{r, j} \sum_{s, k} c_{\beta j}^r c_{\al k}^s
\bra{r\, j} Q \ket{s\, k}.
\eeq
``Mixed'' terms proportional to \eg\ $c_\al c_{\beta j}^r$ are absent by
the assumption eq.\ \assum.
Demanding that the physical states be normalized to unity gives
$c_\al = 1 + O(\lam^2)$, and the result eq.\ \AG\ follows immediately.

Usually, simple current algebra arguments such as this
are spoiled by nonanalyticity in $M_q$ due to the presence of massless
Nambu--Goldstone bosons in the limit $M_q \goto 0$.
However, we note that if we write
\eq
\label\quarkmass
M_q = m_0 1 + \delta m\, T_8,
\eeq
and consider an expansion in $\delta m$ with $m_0$ held fixed, there
are no massless particles, and we expect that physical quantities are
analytic in $\delta m$.
The AG theorem then guarantees that corrections to the vector form factors
are $O(\delta m^2)$.
This is not the limit relevant for the real world, where
$m_0, \delta m \sim m_s$, but we will use this limit to check whether
our calculations are consistent with the AG theorem.

Note that that the AG theorem is not trivially manifest in the effective
lagrangian.
The lagrangian contains terms which appear to give tree-level corrections
to the $f_1$ form factor of order $\delta m$, in violation of the AG
theorem:
\eq
\label\AGviol
\eqalign{
\delta\scr L &= \frac{c_1}\Lam \tr\left(M\right)
\tr\left(\mybar B iv\cdot\nabla B\right)
+ \frac{c_2}{\Lam} \left[\tr\left(\mybar B M iv\cdot\nabla B\right)
+ \hc\right] \cr
& \qquad + \frac{c_3}{\Lam} \left[\tr\left(\mybar B iv\cdot\nabla B M\right)
+ \hc \right]. \cr}
\eeq
(There are other terms which can be related to these by integration by
parts.)
However, these terms also modify the kinetic term for the baryons so
that there is no order $\delta m$ correction to $f_1$.
To see this, we make the field redefinition
\eq
B' = \left[1 + \frac{c_1}{2\Lam} \tr\left(M\right)\right] B
+ \frac{c_2}\Lam M B + \frac{c_3}\Lam B M.
\eeq
$B'$ is a good interpolating field for the baryons provided that $M_q$ does
not mix octet states.
The lagrangian expressed in terms of $B'$ does not contain any terms of the
form eq.\ \AGviol, and the AG theorem is manifest.

\section{Results}

\def\mfunc#1#2{\frac{#1^2 #2^2}{#1^2 - #2^2} \ln\frac{#1^2}{#2^2}}

The one-loop graphs contributing to the vector form factor are shown in
fig.\ 1.
We write
\eq
f_1^{abc}(0) = \alpha_{ab}^c\left(1
+ \frac{1}{16\pi^2f^2}\beta_{ab}^c + \frac{m_K}{16\pi f^2}\gamma_{ab}^c\right),
\eeq
where the well-known lowest-order results are
\eq
\eqalign{
\alpha^{4 + i5}_{p\Lambda} &= -\sqrt{\sfrac 32}, \cr
\alpha^{4 + i5}_{n\Sigma^-} &= -1, \cr
\alpha^{4 + i5}_{\Lambda\Xi^-} &= \sqrt{\sfrac 32}, \cr
\alpha^{4 + i5}_{\Sigma^0 \Xi^-} &= \sfrac 1{\sqrt 2}. \cr}
\eeq
For the $O(m_s)$ corrections, we obtain
\eq
\label\lameq
\eqalign{
\beta_{p\Lambda}^{4 + i5} &= 2 \lambda_1
- D^2 \lam_2 - F (2 D + 3 F) \lam_1, \cr
\beta_{n\Sig^-}^{4 + i5} &= 2 \lambda_1
- D^2 \lam_3 + 3 F (2 D - F) \lam_1, \cr
\beta_{\Lam \Xi^-}^{4 + i5} &= 2 \lambda_1
- D^2 \lam_2 + F (2 D - 3 F) \lam_1, \cr
\beta_{\Sig^0 \Xi^-}^{4 + i5} &= 2 \lambda_1
- D^2 \lam_3 - 3 F (2 D + F) \lam_1, \cr}
\eeq
where
\eq
\eqalign{
\lambda_1 &= \frac 3{16}\biggl(m_\pi^2 + 2 m_K^2 + m_\eta^2
- 2\,\mfunc{m_K}{m_\pi} - 2\,\mfunc{m_\eta}{m_K} \biggr), \cr
\lam_2 &= \frac 1{16}\biggl(9 m_\pi^2 + 10 m_K^2 + m_\eta^2
- 18\,\mfunc{m_K}{m_\pi} - 2\,\mfunc{m_\eta}{m_K} \biggr), \cr
\lam_3 &= \frac 1{16}\biggl(m_\pi^2 + 10 m_K^2 + 9 m_\eta^2
- 2\,\mfunc{m_K}{m_\pi} - 18\,\mfunc{m_\eta}{m_K} \biggr). \cr}
\eeq
The combination of masses defined above are easily seen to satisfy the AG
theorem in the sense discussed in section 3.
Our numerical results are summarized in table 1.
The $O(m_s)$ corrections are $\lsim 5\%$ for all decays, significantly
smaller than what is expected on the basis of dimensional analysis.
We are therefore led to consider the higher order corrections to determine
whether they are numerically important.

The $O(m_s^{3/2})$ contributions proportional to $m_K \Delta_B$ are computed
from the graphs in fig. 1.
Because of the length and unilluminating nature of the resulting formulas,
we will give formulas only for the
case $m_\pi = 0$, and simplify the results using the Gell-Mann--Okubo relations
\eq
\eqalign{
m_\eta^2 &= \sfrac 43 m_K^2, \cr
M_\Xi &= \sfrac 32 M_\Lam + \sfrac 12 M_\Sigma - M_N. \cr}
\eeq
(We have checked that the full expressions satisfy the AG theorem.)
We obtain
\eq
\label\lameq
\eqalign{
\gam_{p\Lambda}^{4 + i5} &= \left[ -\sfrac 1{10}(25 - 16\sqrt 3) D^2
-\sfrac 15 (39 - 16\sqrt 3) DF + \sfrac 3{10} (25 - 16\sqrt 3) F^2
\right] M_n \cr
& \qquad + \sfrac 12 D (D - F) M_\Sig \cr
& \qquad + \left[ \sfrac 25 (5 - 4\sqrt 3) D^2
+ \sfrac 1{10} (83 - 32\sqrt 3) DF - \sfrac 3{10} (25 - 16\sqrt 3) F^2
\right] M_\Lam, \cr
\gam_{n\Sig^-}^{4 + i5} &= \left[ -\sfrac 1{30} (103 - 48\sqrt 3) D^2
- \sfrac 15 (39 - 16\sqrt 3) DF + \sfrac 1{10} (103 - 48\sqrt 3) F^2
\right] M_n \cr
& \qquad + \left[ \sfrac 25 (9 - 4\sqrt 3) D^2
+ \sfrac 1{10} (83 - 32\sqrt 3) DF - \sfrac 1{10} (103 - 48\sqrt 3) F^2
\right] M_\Sig \cr
& \qquad - \sfrac 16 D(D - 3F) M_\Lam, \cr
\gam_{\Lam \Xi^-}^{4 + i5} &= \left[ \sfrac 1{10} (25 - 16\sqrt 3) D^2
- \sfrac 15 (39 - 16\sqrt 3) DF - \sfrac 3{10} (25 - 16\sqrt 3) F^2
\right] M_n \cr
& \qquad + \left[ - \sfrac 1{20} (15 - 16\sqrt 3) D^2
+ \sfrac 25 (11 - 4\sqrt 3) DF + \sfrac 3{20} (25- 16\sqrt 3) F^2
\right] M_\Sig \cr
& \qquad + \left[ - \sfrac 1{20} (35 - 16\sqrt 3) D^2
+ \sfrac 15 (17 - 8 \sqrt 3) DF + \sfrac 3{20} (25 - 16\sqrt 3) F^2
\right] M_\Lam, \cr
\gam_{\Sig^0 \Xi^-}^{4 + i5} &= \left[ \sfrac 1{30} (103 - 48\sqrt 3) D^2
- \sfrac 15 (39 - 16\sqrt 3) DF - \sfrac 1{10} (103 - 48\sqrt 3) F^2
\right] M_n \cr
& \qquad + \left[ \sfrac 1{60}(113 - 48\sqrt 3) D^2
- \sfrac 25 (11 - 4\sqrt 3) DF - \sfrac 1{20} (103 - 48\sqrt 3) F^2
\right] M_\Sig \cr
& \qquad + \left[ \sfrac 1{60} (319 - 144\sqrt 3) D^2
+ \sfrac 15 (61 - 24\sqrt 3) DF + \sfrac 3{20} (103 - 48\sqrt 3) F^2
\right] M_\Lam. \cr}
\eeq
Our numerical results (including $m_\pi \ne 0$) are summarized in
table 1.
The main feature of these results is that they are significantly smaller than
expected from dimensional analysis.
Using the older values of $D$ and $F$ or neglecting the pion mass results in
substantially larger corrections.

\section{Conclusions}

We have computed chiral corrections to the $f_1$ vector form factor for
$\Delta S = 1$ semileptonic hyperon decay.
We have shown that the leading corrections are $O(m_s)$ and $O(m_s^{3/2})$ and
explained how this is consistent with the Ademollo--Gatto theorem.
These corrections are $\sim 30\%$ according to dimensional analysis.
Explicit calculation shows that the $O(m_s)$ corrections and a computable
subset of the $O(m_s^{3/2})$ corrections are $\sim 5\%$ for all decays
using the values of $D$ and $F$ of ref.\ \fit.
The corrections are significantly larger if the older values of $D$ and
$F$ are used, or if the pion mass is neglected.

These results are very different than those obtained from the
non-relativistic quark model and bag model \quarkmod.
In these models, the corrections to $f_1$ are universal for all decays
and are $\simeq -1\%$.
The corrections we have computed are much larger and depend on the decay.
{}From the point of view of the chiral expansion, the quark model results
are rather hard to understand.
Since these model predictions are used in a determination of $V_{us}$ from
semileptonic hyperon decay \determ, this discrepancy is of more
than academic interest.
It is possible that the inclusion of the effects of the decuplet \fit\ may
reduce the apparent descrepancy between the chiral corrections and the
quark model calculations.

\vbox{
\vskip 20pt
\centerline{
\vbox{\offinterlineskip
\hrule
\halign{&\vrule#&\strut\quad\hfil#\quad\cr
height2pt&\omit&&\omit&&\omit&&\omit&&\omit&\cr
&\omit\hfil&&$O(m_s)\ \ \ $&&$O(m_s^{3/2})\ \ \ \ $&&total\ \ \ \ \ &\cr
height2pt&\omit&&\omit&&\omit&&\omit&&\omit&\cr
\noalign{\hrule}
height1pt&\omit&&\omit&&\omit&&\omit&&\omit&\cr
\noalign{\hrule}
height2pt&\omit&&\omit&&\omit&&\omit&&\omit&\cr
&$\Lam\goto p\ \ $&&$-0.001 \pm 0.026$&&$0.047 \pm 0.032$&&
$0.046 \pm 0.016$&\cr
height1pt&\omit&&\omit&&\omit&&\omit&&\omit&\cr
\noalign{\hrule}
height1pt&\omit&&\omit&&\omit&&\omit&&\omit&\cr
&$\Sig^-\goto n\ \ $&&$0.065 \pm 0.011$&&$0.060 \pm 0.041$&&
$0.13 \pm 0.05\;\ $&\cr
height1pt&\omit&&\omit&&\omit&&\omit&&\omit&\cr
\noalign{\hrule}
height1pt&\omit&&\omit&&\omit&&\omit&&\omit&\cr
&$\Xi^-\goto \Lam\ \ $&&$0.021 \pm 0.014$&&$0.060 \pm 0.033$&&
$0.081 \pm 0.020$&\cr
height2pt&\omit&&\omit&&\omit&&\omit&&\omit&\cr
height1pt&\omit&&\omit&&\omit&&\omit&&\omit&\cr
\noalign{\hrule}
height1pt&\omit&&\omit&&\omit&&\omit&&\omit&\cr
&$\Xi^-\goto\Sig^0$&&$-0.002 \pm 0.031$&&$0.035 \pm 0.034$&&
$0.033 \pm 0.013$&\cr
height2pt&\omit&&\omit&&\omit&&\omit&&\omit&\cr}
\hrule}}
{\leftskip=30pt\rightskip=30pt\noindent
Table 1: $f_1 / f_1^{SU(3)}$ for $\Delta S = 1$ hyperon decays using the
best-fit values $D = 0.61$, $F = 0.40$ of ref.\ \fit.
The quoted errors are obtained by (somewhat arbitrarily) assigning a $20\%$
error to the values of $D$ and $F$.
Using the older values $D = 0.8$, $F = 0.5$ \oldfit\ increases all of the
corrections by $\sim 40\%$.
We have kept $m_\pi \ne 0$;
taking $m_\pi = 0$ increases all of the corrections, some by as much as $35\%$.
\par}
\vskip 20pt}

\section{Acknowledgements}

We would like to thank M. Suzuki for helpful discussions and R. N. Cahn for
critical reading of the manuscript.
This work was supported by the Director, Office of Energy Research, Office of
High Energy and Nuclear Physics, Division of High Energy Physics of the
U.S.\ Department of Energy under Contract DE-AC03-76SF00098.


\listrefs
\vfill
\eject
\centerline{\bf Figure Captions}
\vskip .4in
\noindent
Fig.\ 1.
Graphs contributing to the vector form factor at one loop.
The solid lines represent baryons, the dashed lines represent mesons,
and cross indicates an insertion of the vector current.
The wavefunction graph (b) vanishes identically.
Graphs (e) and (f) do not contribute to the $m_K \Delta_B/ 16\pi f^2$
corrections.
\bye